# Temperature-dependent nonlinear phonon shifts in a supported MoS$_2$ monolayer

*Andrzej Taube†, Jarosław Judek, Cezariusz Jastrzębski, Anna Duzynska, Krzysztof Świtkowski and Mariusz Zdrojek\**

Faculty of Physics, Warsaw University of Technology, Koszykowa 75, 00-662 Warsaw, Poland

ABSTRACT

We report Raman spectra measurements on a MoS$_2$ monolayer supported on SiO$_2$ as a function of temperature. Unlike in previous studies, the positions of the two main Raman modes, $E^1_{2g}$ and $A_{1g}$ exhibited nonlinear temperature dependence. Temperature dependence of phonon shifts and widths is explained by optical phonon decay process into two acoustic phonons. Based on Raman measurements local temperature change under laser heating power at different global temperatures is derived. Obtained results contribute to our understanding of the thermal properties of two-dimensional atomic crystals and can help to solve the problem of heat dissipation, which is crucial for use in the next generation of nanoelectronic devices.

**KEYWORDS**
**MoS$_2$ monolayer, two dimensional atomic crystals, Raman spectroscopy, thermal properties, phonons, laser heating**

Two-dimensional atomic crystals, such as graphene, hexagonal boron nitride and transition metal dichalcogenides (TMDCs), have attracted considerable attention because of their unique electrical, optical and mechanical properties[1-6]. The last class of materials is especially interesting for the next generation of electronic and optoelectronic devices because in contrast to graphene, TMDCs are semiconductors with a non-zero bandgap. Among them, molybdenum disulphide, MoS$_2$, has been the most intensively studied. Its 1.85 eV direct bandgap and presumable high mobility[7], are properties desirable for the construction of field effect transistors[8-9], sensors[10] and photodetectors[11] as well as in the emerging field of valleytronics[12]. Knowledge of a material's thermal properties and a interfacial thermal resistance between adjacent layers[13] is critical for the operation of a variety of electronic devices. This is because heat dissipation is currently one of the most significant constraints on the design and fabrication of integrated electronic circuits. Although MoS$_2$ has been investigated extensively, there have been few studies[14,15] on its thermal properties, and to our knowledge, none was strictly devoted to supported monolayers of MoS$_2$. This configuration is important, because it represents the most commonly used device configuration.

In this article, we describe the temperature-dependent nonlinear phonon shifts in supported MoS$_2$ monolayers. To access the thermal properties of single MoS$_2$ layers, we used Raman scattering, which is a simple, convenient and reliable tool used to characterise nanomaterials, such as supported and suspended graphene[16,17] or few-layer MoS$_2$ flakes at room temperature[14,15]. Especially temperature-dependent Raman shifts can be used to investigate vibration, transport, phonon-phonon properties or electron-phonon interactions[18]. In contrast to previous Raman studies, we show that the positions of the two main Raman modes, $E_{2g}^1$ and $A_{1g}$, exhibit nonlinear temperature dependence stemming from optical phonon decay. Based on obtained results we calculate how local temperature of MoS$_2$ monolayer changes with different laser power levels.

A schematic of the experimental setup is shown in FIG. 1a, in which a MoS$_2$ flake placed on the Si/SiO$_2$ substrate is illuminated by a laser beam focused on a small area on the flake. The laser's role is twofold. First, it acts as a source of photons that can be inelastically scattered by phonons in a typical Raman measurement. Second, it serves as a heat source because it can carry a high level of energy that, when absorbed, can significantly increase the local temperature of the sample. Importantly, this temperature increase is reflected in the observed optical phonon energies and lifetimes. Therefore, Raman spectroscopy allows us to increase the local sample temperature and simultaneously yields information about the temperature.

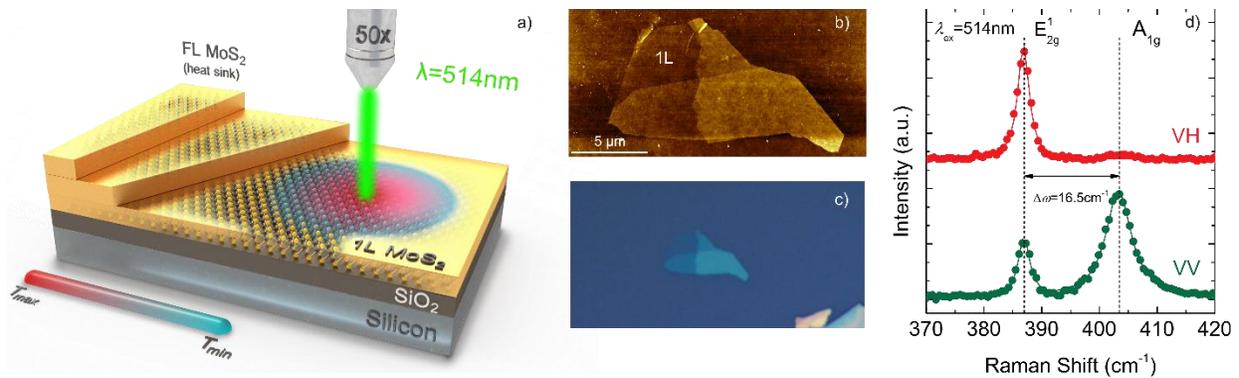

**Figure 1.** (a) Schematic of the experimental setup; (b) AFM picture of a typical MoS$_2$ flake and (c) its image from an optical microscope; (d) Raman spectrum of the MoS$_2$ monolayer at the room temperature and under vacuum for the VV (bottom) and VH (top) light polarization configuration.

First, the dependence of the phonon energies on temperature is carefully measured. The substrate and the entire MoS$_2$ flake are the same temperature, and the power of the laser is maintained low enough to not increase the temperature of the irradiated spot. Subsequently, the phonon energy change for the change in laser power level is measured. Combining these measurements, it is possible to determine how the increase of the power level changes the local temperature. We intentionally placed the MoS$_2$ monolayers on a dielectric substrate because local temperature determined in this way better describes that in real devices.

The MoS$_2$ flakes were fabricated using conventional mechanical exfoliation from MoS$_2$ single crystals (SPI) and then transferred to the Si/SiO$_2$ substrate[19]. Optical microscopy, atomic force microscopy and Raman spectroscopy were employed to identify the monolayers[20]. FIG. 1b shows a folded flake, with the region denoted as 1L confirmed to be a single layer. The thickness from AFM studies was 0.65-0.7 nm, typical for a MoS$_2$ monolayer on a SiO$_2$ substrate[21]. Raman spectra taken by a Dilor XY-800 spectrometer using an Ar laser at 514 nm (2.41 eV) excitation in the VV light polarisation configuration (the electric field vectors of the incident light and scattered light are parallel) show two active modes, denoted[22] as $E_{2g}^1$ and $A_{1g}$ (FIG. 1d). In the VH light polarisation (the electric field vectors of the incident light and scattered light are perpendicular) only $E_{2g}^1$ mode was observed. Therefor VV configuration is more suitable for our experiment. Differences in Raman spectra taken in VV and VH configuration originate in symmetry selection rules.

A single layer of molybdenum disulphide consists of two hexagonal planes of sulphur atoms linked by covalent bonds with the hexagonal plane of molybdenum atoms in the middle. The $A_{1g}$ mode is associated with the out-of-plane vibrations of the sulphur atoms, whereas the $E_{2g}^1$ mode is related to the in-plane vibrations of the sulphur and molybdenum atoms. The difference between the positions of these two modes depends on the number of layers in the flake, and for a monolayer equals 18.5 cm$^{-1}$ under ambient atmosphere[23] and 16.5 cm$^{-1}$ in vacuum. This effect is attributed to environmental effects, particularly to the desorption of various molecules from the MoS$_2$ surface[24]. Temperature-dependent Raman spectra were obtained while heating and cooling the samples in a microscope cryostat in which the temperature was controlled between 70 K and 350 K with a stability of approximately 0.1 K. The laser power on the sample was carefully calibrated and set between 30 and 60 μW.

The results of the temperature-dependent Raman study are presented in FIG. 2. In FIG. 2a, spectra for selected temperatures from 70 K to 350 K are shown to illustrate the Raman spectra evolution. The downshift and broadening of the peaks can be easily observed. However, to perform a more quantitative analysis, we calculated the positions and widths (FWHM) of both observed peaks and plotted them in FIG. 2b and 2c (see ref. 25 for details). Interestingly, at low temperatures it can be seen a clear deviation from linear temperature dependence of both positions and the linewidth. We note that subsequent measurements and measurements on different samples not give much difference in obtained results. We remind that the peak position (Raman shift) is attributed to the phonon energy ($\hbar\omega$), whereas the FWHM is related to the phonon lifetime.

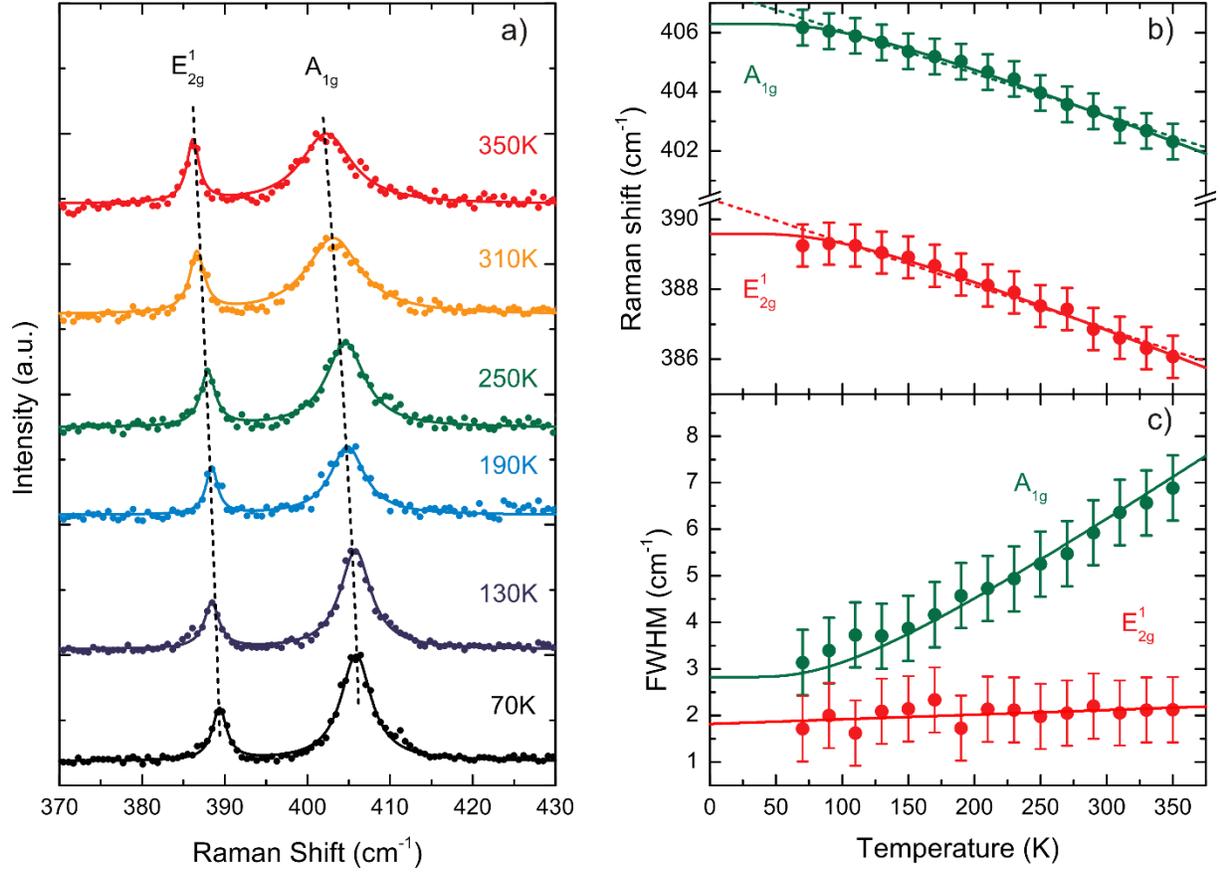

**Figure 2.** a) Raman spectra of MoS$_2$ monolayers measured at selected temperatures from 70 to 350 K. b) Temperature dependence of the Raman peak positions for the $E^1_{2g}$ and $A_{1g}$ modes. The dashed line represents the fit to equation 2, and the solid line is the fit to equation 3. c) Measured values of the phonon line widths at temperatures from 70 to 350 K. The solid line for the $A_{1g}$ mode represents the fit from equation 4. The error bars represent the expanded uncertainty at the 95% level of confidence and include finite spectral resolution and the average uncertainty of determination of peak position (~0.1 cm$^{-1}$) or FWHM (~0.2 cm$^{-1}$).

Generally, temperature dependence of Raman shifts and linewidths arises from electron-phonon, anharmonic phonon-phonon interactions and thermal expansion[18,26]. Particularly, one can use these factors to understand observed temperature dependence of Raman spectra. For example negative temperature dependence of graphene G-mode linewidth after Ar annealing can be associated to electron-phonon interactions[26] whereas positive temperature dependence of multilayer MoS$_2$ A$_{1g}$ mode linewidth can be explained by optical phonon decay[14]. In our case , to describe the phonon softening (decrease in phonon energy) due to the temperature increase, we employed two approaches that have been used in numerous studies. In the first one, a linear change in the energy with temperature is assumed by:[17,27]

$$\omega(T) = \omega_0 + \chi T \qquad (1)$$

where $\omega_0$ is the phonon frequency at zero temperature and $\chi$ is the first-order temperature coefficient. Fitting our results to expression (1), we obtained parameters that were in good agreement with other works (see Table 1 for details). The linear fit to our data is represented by the dashed line in FIG. 2b.

|  | $\chi$ (cm$^{-1}$/K) | |
| --- | --- | --- |
|  | $A_{1g}$ | $E^1_{2g}$ |
| This work, sup. monolayer, EX | -0.0143 | -0.0124 |
| Ref. 14, susp. multilayer, EX | -0.0123 | -0.0132 |
| Ref. 15, sup. monolayer, EX | -0.013 | -0.017 |
| Ref. 15, susp. monolayer, EX | -0.013 | -0.011 |
| Ref. 27, sup. monolayer, CVD | -0.016 | -0.013 |
| Ref. 27, sup. multilayer, EX | -0.013 | -0.015 |

**Table 1.** Temperature dependence of the positions of the $E^1_{2g}$ and $A_{1g}$ MoS$_2$ modes for suspended (susp.) or substrate supported (sup.) layers. CVD: chemical vapor deposition; EX: mechanical exfoliation

The second approach is based on a phenomena of the optical phonon decay of two acoustic phonons with equal energies due to lattice potential anharmonicity. A useful formula was proposed by Balkanski[28]:

$$\omega(T) = \omega_0 + A\left(1 + \frac{2}{e^x - 1}\right) \quad (2)$$

where $x = \hbar(\omega_0/2)/k_B T$ and $A$ is an anharmonic constant. The values of the fitting parameters are listed in Table 2. We found that this approach, represented by the solid lines in FIG. 2b, was more appropriate for the description of $\omega(T)$ because it better fit the experimental data at low temperatures (<150 K). We note that measurable changes in phonon energies versus temperature are a necessary condition for the applicability of our method. Therefore, in the case of saturation of the phonon energies at low $T$, it could be impossible to reliably estimate small temperature changes from the peak position changes. Fortunately, this situation does not take place in our study. It is also interesting that such nonlinear behavior, shown here in the case of single-layer MoS$_2$, was also recently observed for supported WS$_2$ single layers[29]. However, the authors attributed this nonlinearity to experimental artefacts and used eq. 1 to describe the phonon energy shifts with temperature.

|  | $A_{1g}$ | $E_{2g}^1$ |
|---|---|---|
| $\omega_0$ (cm$^{-1}$) | 408.9 | 391.7 |
| $A$ (cm$^{-1}$) | -2.61 | -2.15 |
| $\Gamma$ (cm$^{-1}$) | 29.82 | - |

**Table 2**. Temperature dependence of the positions and FWHMs of the $E_{2g}^1$ and $A_{1g}$ modes. For the $E_{2g}^1$ mode, the changes in the FWHM are lower than the uncertainties.

For higher temperatures, taking into account only first order Taylor expansion, the two theoretical curves converge because equation (2) converts to a linear dependence:

$$\omega(T) \xrightarrow{x=\frac{\hbar\omega_0}{2k_BT}\ll 1} \omega_0 + A + \frac{4Ak_B}{\hbar\omega_0}T = \omega_0' + \chi_B T, \quad \omega_0' = \omega_0 + A, \quad \chi_B = \frac{4Ak_B}{\hbar\omega_0}T \quad (3)$$

with slope values of -0.0178 cm$^{-1}$/K and -0.0152 cm$^{-1}$/K for the $E_{2g}^1$ and $A_{1g}$ modes, respectively. Differences from the data in Table 1 originate from the smaller number of experimental points in the linear regime.

Figure 2c shows the temperature dependence of the Raman peak widths (FWHM) for both the experimental data and the theoretical curves. It can be seen that the changes in the width of the $E_{2g}^1$ peak are almost negligible. The linear fits are only a guide for the eye. On the contrary, the width of the $A_{1g}$ peak exhibited strong temperature changes. The theoretical dependence of the FWHM of the peak is obtained by the expression[28]:

$$\Gamma(T) = \Gamma_0 \left(1 + \frac{2}{e^x - 1}\right) \quad (4)$$

where $\Gamma_0$ is the peak width at zero temperature. The fit parameter is shown in Table 2 only for the $A_{1g}$ mode because the $E_{2g}^1$ mode does not follow the predictions of this theory. The peak width could also be considered a temperature indicator. Particularly one can use $\Delta\Gamma$ instead of $\Delta\hbar\omega$. In our work, we used a traditional approach because the uncertainties in estimating the temperature from the peak position shifts were lower than those from the peak width.

We have shown the experimental data and theoretical models for both Raman modes. However, for further data processing, as a temperature indicator, the position of the $A_{1g}$ symmetry mode was chosen because of its higher intensity and slightly larger changes with $T$. For the temperature estimation, we used the dependence determined by fitting with eqs. (1) and (2) instead of taking the raw data.

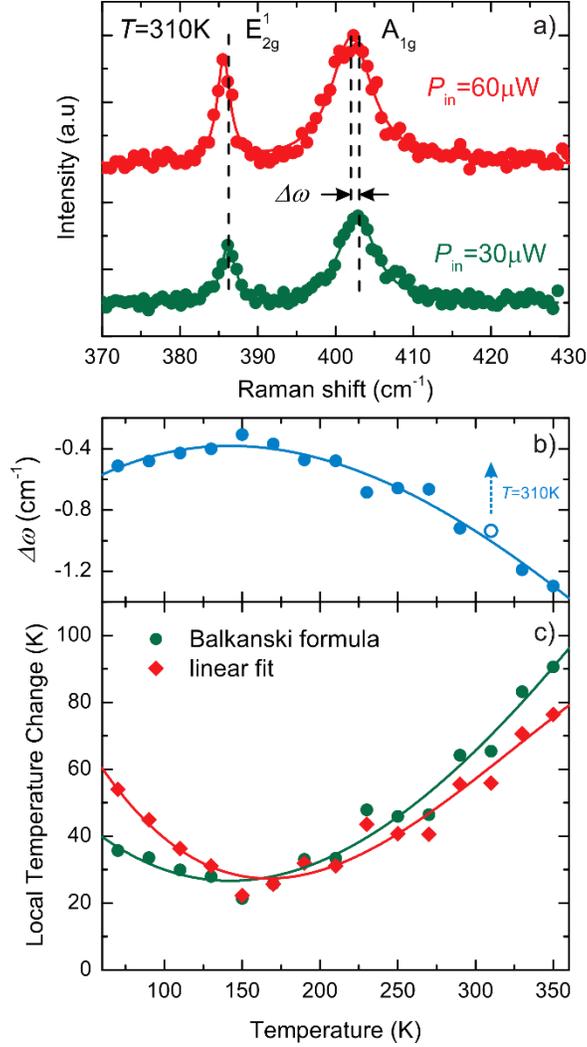

**Figure 3.** a) Raman spectra of single layer MoS$_2$ for different incident laser powers taken at 310 K. b) Change in the $A_{1g}$ peak position due to the absorbed laser power at different temperatures. c) Calculated local temperature change. All solid lines are only guides for viewing.

To determine how laser power level changes local MoS$_2$ temperature, we measured two Raman spectra at every temperature, one for $P_L$=30 µW and the second for $P_L$=60 µW, where $P_L$ represents the laser power on the sample. An example result taken for $T$=310 K is shown in FIG. 3a. For higher laser powers, the Raman peaks clearly downshift as a result of local heating. The complete set of peak frequency changes is presented in FIG. 3b. We note that the dependence of the peak position is not a monotonic function of temperature and has a maximum at $T$=150 K. Below this temperature, the heating process becomes less effective owing to optical phonon decay, as explained above. Calculations for the temperature rise due to the laser power increase were performed in the following way. At a fixed global temperature (temperature of the cryostat and the entire sample without laser illumination), we calculated the temperature corresponding to the $A_{1g}$ mode peak positions for the two laser powers (using the data presented in Figure 2b and 3b). The difference was called the local temperature change and is depicted in FIG. 3c. The minimum $\Delta T$ increase coincides with the minimal Raman shift from FIG. 3b. We note that the local temperature was estimated on the basis of both fitted curves describing the temperature dependence of the Raman peak position for mode $A_{1g}$. The divergences at low and high temperatures are quite intuitive with respect to FIG. 2b and originate in the differences between the assumed models discussed earlier. The minimum of $\Delta T$ indicates that the most efficient heat dissipation in supported MoS$_2$ monolayer occurs at $T$=150K. This knowledge can be used for proper thermal management of nanoscale electrical devices. This is important finding, because in real devices, like graphene transistors on SiO$_2$/Si substrates, most of the heat power is dissipated by layers lying below conductive channel. Local temperature change at different global temperature and constant heating power might be related to changes in the thermal conductivity[16] or the interfacial thermal resistance[13]. Our results may be further used to determine temperature dependence of above mentioned quantities.

In conclusion, we have experimentally studied the temperature-dependent Raman spectra of a supported MoS$_2$ monolayers at temperatures between 70 K and 350 K. We found that Raman shifts exhibits nonlinear temperature dependence. We conclude that obtained effects stem from the optical phonon decay and lattice potential anharmonicity in the MoS$_2$ monolayer. In high temperatures, phonon shifts was described by first order temperature coefficients which was -0.0178 cm$^{-1}$/K and -0.0152 cm$^{-1}$/K for the $A_{1g}$ and $E_{2g}^1$ modes, respectively. Based on Raman measurements local temperature change of supported MoS$_2$ monolayer under laser heating power was calculated as a function of different global temperatures. The minimum temperature rise $\Delta T$=20 was observed at $T$=150 K. We claim that observed effect is related to thermal conductivity changes of monolayer MoS$_2$ and more effort is needed to explain this in details. Our results may be useful for further experimental and theoretical studies on the thermal properties of two-dimensional atomic crystals based on MoS$_2$ and other dichalcogenides single layers.

AUTHOR INFORMATION

**Corresponding Author**

*E-mail: zdrojek@if.pw.edu.pl

**Present Addresses**

†Andrzej Taube is also with Institute of Electron Technology, Warsaw, Poland and with Institute of Microelectronics and Optoelectronics, Warsaw University of Technology, Warsaw, Poland.

**Notes**

The authors declare no competing financial interest.

ACKNOWLEDGMENT

The work was supported by the Polish Ministry of Science within the Diamond Grant programme (0025/DIA/2013/42). AT and AD were also supported by the European Union in the